\documentclass[11pt,a4paper]{article}
\usepackage{amsfonts,amssymb}
\usepackage{cite}
\usepackage[T2A]{fontenc}
\usepackage[cp1251]{inputenc}
\usepackage[english]{babel}
\usepackage{amsmath}%
\usepackage[unicode]{hyperref}

\title{\bf Invariant manifolds and separation of the variables for integrable chains.}
\author{\bf I.T. Habibullin and   A.R. Khakimova}

\date{}

\begin{document}

\maketitle

\begin{abstract}
A notion of the generalized invariant manifold for a nonlinear integrable lattice is considered. Earlier it has been observed that this kind objects provide an effective tool for evaluating the recursion operators and Lax pairs. In this article we show with an example of the Volterra chain that the generalized invariant manifold can be used for constructing exact particular solutions as well. To this end we first find an invariant manifold depending on two constant parameters. Then we assume that ordinary difference equation defining the generalized invariant manifold has a solution polynomially depending on one of the spectral parameters and derive ordinary difference and differential equations, for the roots of the polynomials. Efficiency of the method is approved by some illustrative examples.
\end{abstract}

\maketitle

%\eqnobysec

\section{Introduction}

In the series of our articles \cite{HabKhaPo}-\cite{KhaUMJ18}, we observed that such important objects for integrable nonlinear equations as the Lax pair and the recursion operator can be effectively obtained directly from the linearized equation. To do this, we construct an ordinary differential (or difference) equation that is compatible with the linearized equation for each solution of the nonlinear equation under consideration. We call this ODE a generalized invariant manifold. The aim of the present article is to show that the concept of invariant manifold provides an effective tool for constructing exact solutions for nonlinear equations. As a basic object we take the well known Volterra chain \cite{Volterra}
\begin{equation}\label{volterra}
\dot{u}_n=u_n(u_{n+1}-u_{n-1}),
\end{equation}
where the desired function $u=u_n(t)$ depends on the real $t$ and the integer $n$, and the dot above the letter means the derivative with respect to time $t$.
Volterra chain is one of the well studied models of the integrability theory \cite{Manakov}, \cite{KacvanMoerbeke}, \cite{Novikovbook}. Its finite-gap solutions are constructed in \cite{Veselov}, \cite{Ver}, \cite{Solomon} by using the discrete version of the algebro-geometric approach (see \cite{DubrovinMatveevNovikov}, \cite{Krichever77}). Various classes of exact self-similar solutions to the chain are found and their applications in the theory of two-dimensional quantum gravitation are discussed in \cite{ItsKitaevFokas}, \cite{FokasItsKitaev}. Alternative methods for studying explicit solutions of the integrable lattices are recently suggested in \cite{AdlerShabat1}, \cite{AdlerShabat2}, \cite{Nijhoff1}.

Let's recall some of the necessary definitions that we are going to use below. For a given equation of the Volterra type
\begin{equation}\label{volterratype}
\dot{u}_n=f(u_{n+1}, u_n,u_{n-1})
\end{equation}
the corresponding linearized equation is evaluated as follows
\begin{equation}\label{volterratypelin}
\dot{U}_n=\frac{\partial f}{\partial u_{n+1}}U_{n+1}+ \frac{\partial f}{\partial u_{n}}U_{n}+\frac{\partial f}{\partial u_{n-1}}U_{n-1}.
\end{equation}
According to the scheme proposed in \cite{HabKhaPo}, we add to the pair of equations (\ref{volterratype}), (\ref{volterratypelin}) an ordinary difference equation of the form
\begin{equation}\label{volterratypeGIM}
U_{n+k}=G(U_{n+k-1},U_{n+k-2},...U_{n-k_1}, [u_n]),\quad k\geq1,\quad k_1\geq0
\end{equation}
compatible with the linearized equation (\ref{volterratypelin}) for each solution of the equation (\ref{volterratype}). The notation $[u_n]$ indicates that function $G$ might depend on the variable $u_n$ and its several shifts $u_{n\pm1}$, $u_{n\pm2}$, \dots . We call the surface on the space of the dynamical variables, which is determined by the equation (\ref {volterratypeGIM}), a generalized invariant manifold. Actually function $G$ satisfies the following determining equation
\begin{equation}\label{determiningGIM}
\frac{dG}{dt}-D^k_n\dot{U}_n=0 \quad \mbox{mod}\,((1.2),(1.3),(1.4)). 
\end{equation} 
Let us explain the scheme of searching $G$ from relation (\ref{determiningGIM}). First, we must exclude in (\ref{determiningGIM}) all time-derivatives using  equations (\ref {volterratype}), (\ref {volterratypelin}), and then we exclude the variables $ U_ {n + m} $ for $ m \geq k $, as well as for $ m \leq -k_1-1 $ using the relation
(\ref {volterratypeGIM}). It is possible since we request that the equation (\ref{volterratypeGIM}) can be rewritten in the form solved with respect to the variable $U_{n-k_1}$. After these manipulations we arrive at an equation which should hold identically with respect to the independent dynamical variables 
$$U_{n+k-1}, U_{n+k-2}, ...U_{n-k_1};u_n,u_{n\pm1}, u_{n\pm2}, \dots .$$
Thus, we get an overdetermined system of equations for unknown function $ G $, which is usually effectively solved. Having found $G$ we get a triple of equations (\ref{volterratype}), (\ref{volterratypelin}) and (\ref{volterratypeGIM}) showing that the dynamics due to equations (\ref{volterratype}), (\ref{volterratypelin}) admits a compatible reduction (\ref{volterratypeGIM}). An intriguing fact is that for an appropriately chosen function $G$ the consistency condition of the equations (\ref{volterratypelin}) and (\ref{volterratypeGIM}) implies (\ref{volterratype}). Actually in such a case a pair of the equations (\ref{volterratypelin}), (\ref{volterratypeGIM}) provides a Lax pair for (\ref{volterratype}) (generally nonlinear!). The purpose of the article is to show that this nonlinear Lax pair can be used for constructing explicit solutions of the equation (\ref{volterratype}). 

Since the invariant manifold deals with two sets of dynamical variables defined by the equation in question and the linearized equation, it has two orders. Sometimes the problem of determining orders causes problems. The corresponding relationship between these two orders for the case of equations of the KdV type was established in \cite{ZhangZY}.

Let us briefly discuss the content of the article. In section 2 for the Volterra chain we evaluate the generalized invariant manifold of the order two, depending on two constant parameters. At the end of the section it is shown that system (\ref{Laxt}), (\ref{Laxn}) provides a nonlinear Lax pair for the Volterra chain. In \S 3 we assumed that this Lax pair admits a polynomial in the parameter $\lambda$ solution and observed that this assumption puts a severe restriction on the solution of the nonlinear lattice in the form of the overdetermined system of difference and differential equations. The idea is to solve this overdetermined set of equations and find the desired particular solutions for the nonlinear chain. A scheme of realization of the idea is illustrated in \S4 by taking the Volterra chain as an example.  Here for $m=1$, using an invariant manifold, we obtained a pair of scalar ordinary equations compatible with each other. One of which is differential, and the other is difference. Then we reduced the differential equation to an equation which is solved in terms of the Weierstrass $\wp$ function. It is worth to note that simultaneously the difference equation turns into the addition theorem for $\wp$. In generic case the problem of constructing a solution $u_n(t)$ is reduced to a system of ordinary differential equations (\ref{dubrt3}), (\ref{dubrt4}) with some additional constraint given by a system of $m$ algebraic equations (\ref{Rj}). These equations are shown in enlarged form for $m=2$ (see (\ref{dubrt_meq2}), (\ref{Rj1_meq2}), (\ref{Rj2_meq2})).

\section{Evaluation of the invariant manifold for the Volterra chain.}

In this section we concentrate on the Volterra chain (\ref{volterra}). First we find its linearization which obviously looks like
\begin{equation}\label{volterralin}
\dot{U}_n=u_n(U_{n+1}-U_{n-1})+(u_{n+1}-u_{n-1})U_n.
\end{equation}
The linearized equation needs some simplification, since it contains three extra variables $u_n, u_{n+1}, u_{n-1}$ which are interpreted here as some functional parameters. Let's change the variables in the latter by setting $U_n=u_n(P_{n+1}-P_{n-1})$. This relation is nothing else but the linearization of the formula $\log u_n=p_{n+1}-p_{n-1}$ relating the Volterra chain with the equation $\dot{p}_n=e^{p_{n+1}-p_{n-1}}$. In terms of the new variables the linearized equation looks much simpler
\begin{equation}\label{volterralin2}
\dot{P}_n=u_n(P_{n+1}-P_{n-1}).
\end{equation}

We look for an invariant manifold for the equation (\ref{volterralin2}) in the following form
\begin{equation}\label{volterralinGIM2}
P_{n+1}=F(P_{n},P_{n-1},u_n).
\end{equation}
To do this, we must solve the defining equation
\begin{equation}\label{determiningGIM2}
\frac{d}{dt}F(P_{n},P_{n-1},u_n)-D_n(u_n(P_{n+1}-P_{n-1}))=0 \, \mbox{mod}\,((1.1),(2.2),(2.3)). 
\end{equation} 
In addition to (\ref{volterralinGIM2}), in our further computations, we also need a function that, in a sense, is inverse to (\ref{volterralinGIM2})
\begin{equation}\label{volterralinGIM2-}
P_{n-1}=G(P_{n+1},P_{n},u_n)
\end{equation}
assuming that function $G$ is single-valued.  Then, an equation of the form 
\begin{equation}\label{volterralinGIM2-2}
P_{n-2}=G(P_{n},P_{n-1},u_{n-1})
\end{equation}
also defines an invariant manifold for the equation (\ref{volterralin2}) if the condition 
\begin{equation}\label{determiningGIM2-}
\frac{d}{dt}G(P_{n},P_{n-1},u_{n-1})-D^{-2}_n(u_n(P_{n+1}-P_{n-1}))=0 \, \mbox{mod}\,((1.1),(2.2),(2.6)) 
\end{equation}
is satisfied. 
Let's rewrite equations (\ref{determiningGIM2}) and (\ref{determiningGIM2-}) in a more detailed form
\begin{equation}\label{determiningGIM2df}
\frac{dF}{dP_{n}}P_{n,t}+\frac{dF}{dP_{n-1}}P_{n-1,t}+\frac{dF}{du_{n}}u_{n,t}-u_{n+1}(P_{n+2}-P_{n})=0, 
\end{equation}
\begin{equation}\label{determiningGIM2-2df}
\frac{dG}{dP_{n}}P_{n,t}+\frac{dG}{dP_{n-1}}P_{n-1,t}+\frac{dG}{du_{n-1}}u_{n-1,t}-u_{n-2}(P_{n-1}-P_{n-3})=0, 
\end{equation} 
$\mbox{mod} \, ((1.1),(2.2),(2.3),(2.6))$.
Replace the variables  $u_{n,t}$, $u_{n-1,t}$, $P_{n,t}$, $P_{n-1,t}$, $P_{n+2}$ and $P_{n-3}$ in equations (\ref{determiningGIM2df}) and (\ref{determiningGIM2-2df}) by virtue of the equations (\ref{volterra}), (\ref{volterralin2}), (\ref{volterralinGIM2}) and (\ref{volterralinGIM2-2}). Then we will get
\begin{eqnarray}\label{determiningGIM2df2}
&u_n(F-P_{n-1})\frac{dF}{dP_{n}}-u_{n-1}(G-P_{n})\frac{dF}{dP_{n-1}}\nonumber\\
&\qquad +u_{n}(u_{n+1}-u_{n-1})\frac{dF}{du_{n}}-u_{n+1}(D_n(F)-P_n)=0,
\end{eqnarray}
\begin{eqnarray}\label{determiningGIM2-2df2}
u_n(F-P_{n-1})\frac{dG}{dP_{n}}-u_{n-1}(G-P_{n})\frac{dG}{dP_{n-1}}\nonumber\\
\qquad +u_{n-1}(u_{n}-u_{n-2})\frac{dG}{du_{n-1}}+u_{n-2}(D^{-1}_n(G)-P_{n-1})=0,
\end{eqnarray} 
where 
\begin{eqnarray*}
&D_n(F(P_n,P_{n-1},u_n))=F(F(P_n,P_{n-1},u_n),P_n,u_{n+1}),\\
&D^{-1}_n(G(P_{n},P_{n-1},u_{n-1}))=G(P_{n-1},G(P_{n},P_{n-1},u_{n-1}),u_{n-2}).
\end{eqnarray*}
We differentiate relations (\ref{determiningGIM2df2}) and (\ref{determiningGIM2-2df2}) twice with respect to $u_{n+1}$ and $u_{n-2}$, respectively. As a result, we obtain 
\begin{eqnarray}\label{determiningGIM2df2_up1}
u_{n+1}\frac{\partial^2}{\partial u^2_{n+1}}D_n(F)+2\frac{\partial}{\partial u_{n+1}}D_n(F)=0,
\end{eqnarray}
\begin{eqnarray}\label{determiningGIM2-2df2_um2}
u_{n-2}\frac{\partial^2}{\partial u^2_{n-2}}D^{-1}_n(G)+2\frac{\partial}{\partial u_{n-2}}D^{-1}_n(G)=0.
\end{eqnarray} 
We apply the shift operator $D^{-1}_n$ to both sides of the equality (\ref{determiningGIM2df2_up1}) and, respectively, the operator $D_n$ to both sides of (\ref{determiningGIM2-2df2_um2}):
\begin{eqnarray}\label{determiningGIM2df2_u}
u_{n}\frac{\partial^2}{\partial u^2_{n}}F+2\frac{\partial}{\partial u_{n}}F=0,
\end{eqnarray}
\begin{eqnarray}\label{determiningGIM2-2df2_um1}
u_{n-1}\frac{\partial^2}{\partial u^2_{n-1}}G+2\frac{\partial}{\partial u_{n-1}}G=0.
\end{eqnarray}
The resulting equations are easily solved
\begin{eqnarray}\label{GIM_F}
F(P_n,P_{n-1},u_n)=F_1(P_n,P_{n-1})+\frac{1}{u_n}F_2(P_n,P_{n-1}),
\end{eqnarray}
\begin{eqnarray}\label{GIM_G}
G(P_n,P_{n-1},u_{n-1})=G_1(P_n,P_{n-1})+\frac{1}{u_{n-1}}G_2(P_n,P_{n-1}).
\end{eqnarray}
We substitute the found expressions into the equations (\ref{determiningGIM2df2}) and (\ref{determiningGIM2-2df2}). Now, since the dependence of the functions $F$ and $G$ on the variables $u_n$ and $u_{n-1}$ is already determined, we can compare in (\ref{determiningGIM2df2}) and (\ref{determiningGIM2-2df2}) the coefficients in front of the powers of the independent variables $u_{n+1}$, $u_{n-1}$ and $u_{n-2}$, $u_{n}$, respectively. As a result, we get a set of equations
\begin{eqnarray}
&\left(P_n-G_1(P_n,P_{n-1})\right)\frac{\partial F_1(P_n,P_{n-1})}{\partial P_{n-1}}=0,\label{GIM_F_eq1}\\
&u_n\left(D_n(F_1(P_n,P_{n-1}))-P_n\right)+F_2(P_n,P_{n-1})=0,\label{GIM_F_eq2}\\
&\left(P_n-G_1(P_n,P_{n-1})\right)\frac{\partial F_2(P_n,P_{n-1})}{\partial P_{n-1}}+F_2(P_n,P_{n-1})=0,\label{GIM_F_eq3}
\end{eqnarray}
\begin{eqnarray}
&\left(F_1(P_n,P_{n-1})-P_{n-1}\right)\frac{\partial G_1(P_n,P_{n-1})}{\partial P_{n}}=0,\label{GIM_G_eq1}\\
&u_{n-1}\left(D^{-1}_n(G_1(P_n,P_{n-1}))-P_{n-1}\right)+G_2(P_n,P_{n-1})=0,\label{GIM_G_eq2}\\
&\left(F_1(P_n,P_{n-1})-P_{n-1}\right)\frac{\partial G_2(P_n,P_{n-1})}{\partial P_{n}}-G_2(P_n,P_{n-1})=0.\label{GIM_G_eq3}
\end{eqnarray}
The equation (\ref{GIM_F_eq1}) confirms that there are two possibilities:
\begin{eqnarray*}
i) \quad &G_1(P_n,P_{n-1})=P_n,\\
ii) \quad &\frac{\partial F_1(P_n,P_{n-1})}{\partial P_{n-1}}=0.
\end{eqnarray*}
The first case leads to a trivial solution
\begin{eqnarray*}
F(P_n,P_{n-1},u_n)\equiv P_{n-1}, \quad G(P_n,P_{n+1},u_{n})\equiv P_{n+1}.
\end{eqnarray*}
Let's focus on $ ii) $, it gives right away
\begin{eqnarray*}
F_1(P_n,P_{n-1})=F_3(P_n).
\end{eqnarray*}
By virtue of the latter and due to the equation (\ref {GIM_G_eq1}) we obtain
\begin{eqnarray*}
G_1(P_n,P_{n-1})=G_3(P_{n-1}).
\end{eqnarray*}
We rewrite equations (\ref{GIM_F_eq2}) and (\ref{GIM_G_eq2}), taking into account the found functions $F_1(P_n,P_{n-1})$ and $G_1(P_n,P_{n-1})$
\begin{eqnarray}
&u_nF_3\left(F_3(P_n)+\frac{1}{u_n}F_2(P_n,P_{n-1})\right)-u_nP_n+F_2(P_n,P_{n-1})=0,\label{GIM_F_eq22}\\
&u_{n-1}G_3\left(G_3(P_{n-1})+\frac{1}{u_{n-1}}G_2(P_n,P_{n-1})\right)-u_{n-1}P_{n-1}\nonumber\\
& \qquad{} \qquad \qquad \qquad{} \qquad{}+G_2(P_n,P_{n-1})=0,\label{GIM_G_eq22}
\end{eqnarray}
We differentiate equation (\ref{GIM_F_eq22}) with respect to $u_n$ and equation (\ref{GIM_G_eq22}) with respect to $u_{n-1}$ twice, then we find that the functions $F_3(P_n)$ and $G_3(P_{n-1})$ have the form
\begin{eqnarray*}
&F_3(P_n)=C_2P_n+C_1,\\
&G_3(P_{n-1})=C_4P_{n-1}+C_3,
\end{eqnarray*}
where $C_i$, $i=1,2,3,4$ are arbitrary constants.
By substituting the found representations of the functions $F_3(P_n)$ and $G_3(P_{n-1})$ into equations (\ref{GIM_F_eq22}) and (\ref{GIM_G_eq22}) and comparing the coefficients before the  variables $u_n$ and $u_{n-1}$, respectively, we find:
\begin{eqnarray*}
&C_2=-1,\quad C_4=-1.
\end{eqnarray*}
Now we substitute the above refinements into the equations (\ref {GIM_F_eq3}) and (\ref {GIM_G_eq3}), and then derive the equations for determining the functions $ F_2 (P_n, P_ {n-1}) $ and $ G_2 (P_n, P_{n-1}) $, which are easily solved:
\begin{eqnarray*}
&F_2(P_n,P_{n-1})=\frac{F_4(P_n)}{C_3-P_n-P_{n-1}},\\
&G_2(P_n,P_{n-1})=\frac{G_4(P_{n-1})}{C_1-P_n-P_{n-1}}.
\end{eqnarray*}
Summing up the above calculations, we can represent the functions $F(P_n,P_{n-1},u_n)$ and $G(P_n,P_{n-1},u_n)$ in the following form:
\begin{eqnarray}
&P_{n+1}=F(P_n,P_{n-1},u_n)=C_1-P_n+\frac{F_4(P_n)}{u_n\left(C_3-P_n-P_{n-1}\right)},\label{GIM_F_4}\\
&P_{n-2}=G(P_n,P_{n-1},u_{n-1})=C_3-P_{n-1}+\frac{G_4(P_{n-1})}{u_{n-1}\left(C_1-P_n-P_{n-1}\right)}.\label{GIM_G_4}
\end{eqnarray}
Let us apply the shift operator $D_n$ to both sides of (\ref{GIM_G_4}) and then express the variable $P_{n+1}$ from the obtained equation:
\begin{eqnarray}
P_{n+1}=F(P_n,P_{n-1},u_n)=C_1-P_n+\frac{G_4(P_n)}{u_n\left(C_3-P_n-P_{n-1}\right)}.\label{GIM_G_41}
\end{eqnarray}
By comparing equations (\ref{GIM_F_4}) and (\ref{GIM_G_41}) we observe that
\begin{eqnarray*}
G_4(P_n)=F_4(P_n).
\end{eqnarray*}
Thus, it remains to determine the only function $F_4(P_n)$.

It is easily observed that the Volterra equation and its linearization are invariant under the reflection transformation $n\rightarrow -n$, $t\rightarrow -t$, therefore it is natural to assume that formulas (\ref{volterralinGIM2}) and (\ref{volterralinGIM2-2}) are related to each other by the reflection transformation as well. Then it follows from formulas (\ref{GIM_F_4}) and (\ref{GIM_G_4}) that $C_1=C_3$. 

Since the equation (\ref{volterralin2}) is invariant under the shift transformation we can remove the parameter $C_1$ by changing 
$P_n=\frac{C_1}{2}+\tilde{P}_n$. Moreover, we are only interested in parameters that cannot be removed, so we put $C_1=0$. 

We return to the equation (\ref{determiningGIM2df2}) and specify it using the obtained formulas: 
\begin{eqnarray}\label{EQ_last}
\left[F_4\left(-P_n-\frac{F_4(P_n)}{u_n\left(P_n+P_{n-1}\right)}\right)-F_4(P_n)\right]u^2_n-\frac{F_4(P_n)F'_4(P_n)u_n}{P_n+P_{n-1}}\nonumber\\
\quad -\frac{(F_4(P_n))^2\left[F'_4(P_n)(P_n+P_{n-1})+F_4(P_{n-1})-F_4(P_n)\right]}{(P_n+P_{n-1})^4}=0.
\end{eqnarray}
Let us rewrite equation (\ref{EQ_last}) in a short form 
\begin{eqnarray}\label{intermediate}
F_4(-x-\delta y)-F_4(x)=c\delta^2+d\delta,
\end{eqnarray}
where $x=P_n$, $\delta=\frac{1}{u_n}$ and so on. Then by taking $\delta=0$ we get that $F_4$ is an even function. 
Differentiation of the equation with respect to $\delta$ three times implies (we can do that because equation (\ref{EQ_last}) is satisfied identically with respect to the variable $u_n$) $F_4'''=0$ which gives 
\begin{eqnarray}\label{F4}
F_4(P_n)=C_5P^2_n+C_6.
\end{eqnarray}
Thus, an invariant manifold (\ref{volterralinGIM2}) is given by an equation of the form
\begin{eqnarray*}
P_{n+1}=-P_n+\frac{\lambda P_n^2+c}{u_n(P_{n}+P_{n-1})},
\end{eqnarray*}
where $\lambda=-C_5$, $c=-C_6$.

Let us summarize the reasonings and computations of this section as a statement.

{\bf Proposition 1}. A system of the equations 
\begin{eqnarray}\label{Laxt}
\dot{P}_n=u_n(P_{n+1}-P_{n-1}),\\
u_n(P_{n+1}+P_n)(P_{n}+P_{n-1})=\lambda P_n^2+c
\label{Laxn}
\end{eqnarray} 
is compatible if and only if the coefficient $u_n=u_n(t)$ solves the Volterra lattice (\ref{volterra}).

We have already proved above that for arbitrary solution $u_n(t)$ to the Volterra chain the system is compatible. The converse can be easily approved by a direct computation. We notice that in fact the system defines a nonlinear Lax pair with two arbitrary constant parameters $\lambda$ and $c$ for the Volterra chain. By applying the operator $D_n-1$ to the equation (\ref{Laxn}) we immediately obtain a linear equation (see \cite{HabKhaPo})
\begin{equation}\label{Recursion}
u_{n+1}(P_{n+2}+P_{n+1})-u_{n}(P_{n}+P_{n-1})=\lambda (P_{n+1}-P_{n})
\end{equation}
which also defines a generalized invariant manifold since it is compatible, as it is easily checked, with the equations (\ref{volterra}), (\ref{volterralin2}). Earlier in \cite{HabKhaTMP18}, we observed that equation (\ref{Recursion}) can be easily rewritten as the recursion operator for the Volterra chain.

\section{Separation of the variables.}

It is easily proved that nonlinear difference equation (\ref{Laxn}), where the constant parameter $c=c(\lambda)$ is in the form
\begin{equation}\label{a1}
c(\lambda)=-\lambda+c^{(0)} +c^{(1)}\lambda^{-1}+c^{(2)}\lambda^{-2}+\dots
\end{equation} 
admits a solution given by the following formal asymptotic expansion:
\begin{equation}\label{a2}
P_n(\lambda)=1+\alpha_n^{(1)}\lambda^{-1}+\alpha_n^{(2)}\lambda^{-2}+\dots.
\end{equation}
Here the coefficients $\alpha_n^{(j)}$ are functions of a finite set of the dynamical variables $u_{n},\,u_{n\pm1},\,u_{n\pm2},\dots$ that are found successively from equation  (\ref{Laxn}). Let us give exact representation for the first two coefficients:
\begin{eqnarray}\label{a3}
\alpha_n^{(1)}&=&2u_n-\frac{1}{2}c^{(0)},\\
\alpha_n^{(2)}&=&\frac{u_n}{2}(\alpha_{n+1}^{(1)}+2\alpha_n^{(1)}+\alpha_{n-1}^{(1)})-\frac{(\alpha_n^{(1)})^2+c^{(1)}}{4}.
\label{a4}
\end{eqnarray} 
If we assume that formal series (\ref{a2}) is terminated at some natural $m$, such that $\alpha_n^{(j)}\equiv 0$ for $\forall j>m$, then we get a system of difference equations for the variables $\alpha_n^{(1)},\,\alpha_n^{(2)},\,\dots,\alpha_n^{(m)}, u_n$. This fact can be used for describing some particular solutions of the Volterra chain. For the terminated case we change slightly our ansatz  (\ref{a1}), (\ref{a2}) by multiplying the given ones by powers of $\lambda$. We rewrite the sought function $P_n(\lambda)$ and function $c(\lambda)$ as polynomials on $\lambda$ and parametrize these polynomials by their roots (earlier this kind of parametrization was used, for instance, in \cite{DubrovinMatveevNovikov}, \cite {Calogero}). That allows us to make the mentioned system of difference equations more symmetrical. Therefore we can  assume that in the equations (\ref{Laxt}), (\ref{Laxn}) the constant parameter $c$ and a solution $P_n$ are polynomials on the parameter $\lambda$ so that
\begin{equation}\label{parametrization}
P_{n}(\lambda)=\prod_{i=1}^{m}(\lambda-\gamma^i_n),\quad c(\lambda)=-\prod_{i=1}^{2m+1}(\lambda-e_i).
\end{equation}
By requesting that $P_n(\lambda)$ is a polynomial we impose to the Lax pair (\ref{Laxt}), (\ref{Laxn}) a constraint of the form  
\begin{equation}\label{polynom}
\frac{\partial^{m+1}}{\partial \lambda^{m+1}}P_n(\lambda)=0.
\end{equation}

{\bf Proposition 2.} Let us suppose that a solution $u_n(t)$ to the Volterra chain (\ref{volterra}) satisfies condition (\ref{polynom})
for $\forall \, n\in \bf{Z}$, $t=0$. Then (\ref{polynom}) is preserved for $t>0$. 

{\bf Proof}. Indeed, the relation (\ref{polynom}) is evidently compatible with equation (\ref{Laxt}).

We substitute the ansatz (\ref{parametrization}) into the equations (\ref{Laxt}), (\ref{Laxn}) and obtain
\begin{eqnarray}\label{paramLaxt}
\sum_{k=1}^{m}\dot{\gamma}^k_n\prod_{i\neq k}(\lambda-\gamma^i_n)&=&u_n\left(\prod_{i=1}^{m}(\lambda-\gamma^i_{n-1})-\prod_{i=1}^{m}(\lambda-\gamma^i_{n+1})\right),
\end{eqnarray}
\begin{eqnarray}
u_n\left(\prod_{i=1}^{m}(\lambda-\gamma^i_{n+1})+\prod_{i=1}^{m}(\lambda-\gamma^i_n)\right)\left(\prod_{i=1}^{m}(\lambda-\gamma^i_n)+\prod_{i=1}^{m}(\lambda-\gamma^i_{n-1})\right)\nonumber\\
=\lambda\prod_{i=1}^{m}(\lambda-\gamma^i_n)^2-\prod_{i=1}^{2m+1}(\lambda-e_i).\label{paramLaxn} 
\end{eqnarray}
By comparing the coefficients in front of the power $\lambda^{2m}$ in (\ref{paramLaxn}) we derive a relation between the field variable $u_n$ and the roots of the polynomial $P_n(\lambda)$
\begin{equation}\label{un}
u_{n}=-\frac{1}{2}\sum_{i=1}^{m}\gamma^i_n+\frac{1}{4}\sum_{i=1}^{2m+1}e_i.
\end{equation}
By setting $\lambda=\gamma^j_n$ and $\lambda=\gamma^j_{n-1}$ for $j=1,2,...,m$ in the equation (\ref{paramLaxn}) we obtain a system of difference equations
\begin{equation}\label{dubrn}
u_n \prod_{i=1}^{m}(\gamma^j_n-\gamma^i_{n+1})\prod_{i=1}^{m}(\gamma^j_n-\gamma^i_{n-1})=-\prod_{i=1}^{2m+1}(\gamma^j_n-e_i),
\end{equation}
\begin{eqnarray}\label{dubrn2}
u_n\left(\prod_{i=1}^{m}(\gamma^j_{n-1}-\gamma^i_{n+1})+\prod_{i=1}^{m}(\gamma^j_{n-1}-\gamma^i_{n})\right)\prod_{i=1}^{m}(\gamma^j_{n-1}-\gamma^i_{n})=\nonumber\\
\qquad \qquad \qquad\gamma^j_{n-1}\prod_{i=1}^{m}(\gamma^j_{n-1}-\gamma^i_{n})^2-\prod_{i=1}^{2m+1}(\gamma^j_{n-1}-e_i).
\end{eqnarray}
Similarly by taking $\lambda=\gamma^j_n$ in (\ref{paramLaxt}) we find 
\begin{equation}\label{dubrt1}
\dot{\gamma}^j_n=\frac{u_n}{\prod_{i\neq j}(\gamma^j_n-\gamma^i_n)}
\left(\prod_{i=1}^{m}(\gamma^j_n-\gamma^i_{n-1})-\prod_{i=1}^{m}(\gamma^j_n-\gamma^i_{n+1})\right)
\end{equation}
by shifting $n\longmapsto n-1$ in (\ref{dubrt1}) we get
\begin{equation}\label{dubrt2}
\dot{\gamma}^j_{n-1}=\frac{u_{n-1}}{\prod_{i\neq j}(\gamma^j_{n-1}-\gamma^i_{n-1})}
\left(\prod_{i=1}^{m}(\gamma^j_{n-1}-\gamma^i_{n-2})-\prod_{i=1}^{m}(\gamma^j_{n-1}-\gamma^i_{n})\right).
\end{equation}
Due to (\ref{dubrn}) we can rewrite (\ref{dubrt1}), (\ref{dubrt2}) as a closed system of ordinary differential equations of the order $2m$
\begin{equation}\label{dubrt3}
\dot{\gamma}^j_n=\frac{u_n}{\prod_{i\neq j}(\gamma^j_n-\gamma^i_n)}
\left(\prod_{i=1}^{m}(\gamma^j_n-\gamma^i_{n-1})+\frac{\prod_{i=1}^{2m+1}(\gamma^j_{n}-e_i)}{u_n\prod_{i=1}^{m}(\gamma^j_n-\gamma^i_{n-1})}\right),
\end{equation}
\begin{equation}\label{dubrt4}
\dot{\gamma}^j_{n-1}=\frac{-u_{n-1}}{\prod_{i\neq j}(\gamma^j_{n-1}-\gamma^i_{n-1})}
\left(\prod_{i=1}^{m}(\gamma^j_{n-1}-\gamma^i_{n})+\frac{\prod_{i=1}^{2m+1}(\gamma^j_{n-1}-e_i)}{u_{n-1}\prod_{i=1}^{m}(\gamma^j_{n-1}-\gamma^i_{n})}\right).
\end{equation}

Let us first exclude the variables $\gamma^i_{n+1}, \, i=1,2,\dots,m$ from the system of discrete equations (\ref{dubrn}), (\ref{dubrn2}). To this end we rewrite equation (\ref{dubrn}) in the form
\begin{equation}\label{Prodeqrj}
\prod_{i=1}^{m}(\gamma^j_{n}-\gamma^i_{n+1})=r^j, \, j=1,2,\dots,m,
\end{equation}
where
\begin{equation*}
r^j=-\frac{\prod_{i=1}^{2m+1}(\gamma^j_{n-1}-e_i)}{u_n\prod_{i=1}^{m}(\gamma^j_n-\gamma^i_{n-1})}.
\end{equation*}
After opening the parentheses in (\ref{Prodeqrj}) we arrive at a system of linear equations 
\begin{equation}\label{lin_eq_gamma_rj}
\alpha^{(1)}(\gamma^j_n)^{m-1}+\alpha^{(2)}(\gamma^j_n)^{m-2}+\dots+\alpha^{(m)}=r^j-(\gamma^j_n)^{m},
\end{equation}
where $j=1,2,\dots,m$ and 
\begin{eqnarray}\label{alpha_j}
\alpha^{(1)}=-\sum^{m}_{i=1}\gamma^{i}_{n+1},\nonumber\\
\alpha^{(2)}=-\sum_{i\neq j}\gamma^{i}_{n+1}\gamma^{j}_{n+1},\\
\dots\nonumber\\
\alpha^{(m)}=(-1)^m\gamma^{1}_{n+1}\gamma^{2}_{n+1}\dots\gamma^{m}_{n+1} \nonumber
\end{eqnarray}
are coefficients of the polynomial $P_{n+1}(\lambda)=\Pi^{i=1}_{m}\left(\lambda-\gamma^{j}_{n+1}\right)$. Equations (\ref{alpha_j}) give explicit representations of the coefficients $\alpha^j$ in terms of $\gamma^1_{n+1}, \gamma^2_{n+1},...\gamma^m_{n+1}$. On the other hand side by solving the linear system (\ref{lin_eq_gamma_rj}) we find representations for the same coefficients in terms of the variables $\gamma^{1}_{n},\, \gamma^{2}_{n},\, \dots, \,\gamma^{m}_{n};$  $\gamma^{1}_{n-1},\, \gamma^{2}_{n-1},\, \dots, \,\gamma^{m}_{n-1}:$ 
\begin{equation}\label{other}
\alpha^j=H^j(\gamma^{1}_{n},\, \gamma^{2}_{n},\, \dots, \,\gamma^{m}_{n};  \gamma^{1}_{n-1},\, \gamma^{2}_{n-1},\, \dots, \,\gamma^{m}_{n-1}).
\end{equation}
By comparing two representations (\ref{alpha_j}), (\ref{other}) we obtain an implicit formula determining dynamics on $n$ for the roots $\gamma^j$.

Now we concentrate on the system (\ref{dubrn2}). Evidently it can be represented in the form
\begin{equation}\label{Prodeqsj}
\prod_{i=1}^{m}(\gamma^j_{n-1}-\gamma^i_{n+1})=s^j, 
\end{equation}
where
\begin{equation*}
s^j=-\prod_{i=1}^{m}(\gamma^j_{n-1}-\gamma^i_{n})+\frac{\gamma^j_{n-1}\prod_{i=1}^{m}(\gamma^j_{n-1}-\gamma^i_{n})^2-\prod_{i=1}^{2m+1}(\gamma^j_{n-1}-e_i)}{u_n\prod_{i=1}^{m}(\gamma^j_{n-1}-\gamma^i_{n})}.
\end{equation*}
From (\ref{Prodeqsj}) we get 
\begin{equation}\label{lin_eq_gamma_sj}
\alpha^{(1)}(\gamma^j_{n-1})^{m-1}+\alpha^{(2)}(\gamma^j_{n-1})^{m-2}+\dots+\alpha^{(m)}=s^j-(\gamma^j_{n-1})^{m}.
\end{equation}
By substituting the earlier found expressions (\ref{other}) for $\alpha^{(1)},\, \alpha^{(2)}\, \dots, \alpha^{(m)}$ into the system (\ref{lin_eq_gamma_sj}) we obtain exactly $m$ constraints relating variables $\gamma^{1}_{n},\, \gamma^{2}_{n},\, \dots, \,\gamma^{m}_{n}$ with the variables $\gamma^{1}_{n-1},\, \gamma^{2}_{n-1},\, \dots, \,\gamma^{m}_{n-1}:$
\begin{equation}\label{Rj}
R^j\left(\gamma^{1}_{n},\, \gamma^{2}_{n},\, \dots, \,\gamma^{m}_{n};\gamma^{1}_{n-1},\, \gamma^{2}_{n-1},\, \dots, \,\gamma^{m}_{n-1}\right)=0,
\end{equation}
$j=1,2,\dots,m.$

Now the problem is to find solution of the finite system of ordinary differential equations (\ref{dubrt3}), (\ref{dubrt4}), satisfying the additional constraint (\ref{Rj}).

Let us consider the system (\ref{dubrt3}), (\ref{dubrt4}) and constraint (\ref{Rj}) for the case $m=2$. For the simplicity we take 
\begin{equation}\label{Pn_meq2}
P_n(\lambda)=(\lambda-\gamma_n)(\lambda-\delta_n).
\end{equation}
Then (\ref{un}) yeilds 
\begin{equation*}
u_n=-\frac{1}{2}(\gamma_n+\delta_n)+\Sigma^{5}_{i=1}e_i. 
\end{equation*}
Functions $\gamma_n=\gamma_n(t), \, \delta_n=\delta_n(t)$ satisfy the following system of differential equations
\begin{eqnarray}\label{dubrt_meq2}
\dot{\gamma}_{n-1}=\frac{-1}{\gamma_{n-1}-\delta_{n-1}}
\left(u_{n-1}(\gamma_{n-1}-\gamma_{n})(\gamma_{n-1}-\delta_{n})+\frac{\prod_{i=1}^{5}(\gamma_{n-1}-e_i)}{(\gamma_{n-1}-\gamma_{n})(\gamma_{n-1}-\delta_{n})}\right),\nonumber\\
\dot{\delta}_{n-1}=\frac{-1}{\delta_{n-1}-\gamma_{n-1}}
\left(u_{n-1}(\delta_{n-1}-\gamma_{n})(\delta_{n-1}-\delta_{n})+\frac{\prod_{i=1}^{5}(\delta_{n-1}-e_i)}{(\delta_{n-1}-\gamma_{n})(\delta_{n-1}-\delta_{n})}\right),\nonumber\\
\dot{\gamma}_n=\frac{1}{\gamma_n-\delta_n}
\left(u_n(\gamma_n-\gamma_{n-1})(\gamma_n-\delta_{n-1})+\frac{\prod_{i=1}^{5}(\gamma_{n}-e_i)}{(\gamma_n-\gamma_{n-1})(\gamma_n-\delta_{n-1})}\right),\\
\dot{\delta}_n=\frac{1}{\delta_n-\gamma_n}
\left(u_n(\delta_n-\gamma_{n-1})(\delta_n-\delta_{n-1})+\frac{\prod_{i=1}^{5}(\delta_{n}-e_i)}{(\delta_n-\gamma_{n-1})(\delta_n-\delta_{n-1})}\right).\nonumber
\end{eqnarray}
Constraints (\ref{Rj}) in this case take the form
\begin{eqnarray}\label{Rj1_meq2}
\frac{\prod_{i=1}^{5}(\gamma_{n-1}-e_i)}{(\gamma_{n-1}-\gamma_{n})^2(\gamma_{n-1}-\delta_{n})^2}+\frac{\prod_{i=1}^{5}(\gamma_{n}-e_i)}{(\gamma_{n-1}-\gamma_{n})^2(\gamma_n-\delta_{n})(\gamma_n-\delta_{n-1})}+\nonumber\\
\qquad \qquad \qquad \frac{\prod_{i=1}^{5}(\delta_{n}-e_i)}{(\gamma_{n-1}-\delta_{n})^2(\delta_{n}-\gamma_n)(\delta_n-\delta_{n-1})}=-2u_n+\gamma_{n-1},
\end{eqnarray}
\begin{eqnarray}\label{Rj2_meq2}
\frac{\prod_{i=1}^{5}(\delta_{n-1}-e_i)}{(\delta_{n-1}-\gamma_{n})^2(\delta_{n-1}-\delta_{n})^2}+\frac{\prod_{i=1}^{5}(\gamma_{n}-e_i)}{(\gamma_{n}-\delta_{n-1})^2(\gamma_n-\delta_{n})(\gamma_n-\gamma_{n-1})}+\nonumber\\
\qquad \qquad \qquad \frac{\prod_{i=1}^{5}(\delta_{n}-e_i)}{(\delta_{n}-\delta_{n-1})^2(\delta_{n}-\gamma_n)(\delta_n-\gamma_{n-1})}=-2u_n+\delta_{n-1}.
\end{eqnarray}
{\bf Proposition 3.} Solution of the system (\ref{dubrt_meq2}) satisfy the following constraint
\begin{eqnarray}\label{PrY}
\frac{\dot{\gamma}_{n}}{(\gamma_{n}-\gamma_{n-1})(\gamma_{n}-\delta_{n-1})}+\frac{\dot{\delta}_{n}}{(\delta_{n}-\gamma_{n-1})(\delta_{n}-\delta_{n-1})}-\nonumber\\
\quad \frac{\dot{\gamma}_{n-1}}{(\gamma_{n-1}-\gamma_{n})(\gamma_{n-1}-\delta_{n})}-\frac{\dot{\delta}_{n-1}}{(\delta_{n-1}-\gamma_{n})(\delta_{n-1}-\delta_{n})}-1=0.
\end{eqnarray}
Proposition 3 is easily proved by using constraints (\ref{Rj1_meq2}), (\ref{Rj2_meq2}) and system (\ref{dubrt_meq2}).

Dynamics on $n$ of the functions $\gamma$, $\delta$ is determined by the system of discrete equations 
\begin{eqnarray}\label{Pr_sump1}
\gamma_{n+1}+\delta_{n+1}=\gamma_{n}+\delta_{n}-\nonumber\\
\frac{1}{u_n(\delta_{n}-\gamma_{n})}\left(\frac{\prod_{i=1}^{5}(\gamma_{n}-e_i)}{(\gamma_n-\gamma_{n-1})(\gamma_n-\delta_{n-1})}-\frac{\prod_{i=1}^{5}(\delta_{n}-e_i)}{(\delta_n-\gamma_{n-1})(\delta_n-\delta_{n-1})}\right),
\end{eqnarray}
\begin{eqnarray}\label{Pr_prp1}
\gamma_{n+1}\delta_{n+1}=\gamma_{n}\delta_{n}-\nonumber\\
\frac{1}{u_n(\delta_{n}-\gamma_{n})}\left(\frac{\delta_{n}\prod_{i=1}^{5}(\gamma_{n}-e_i)}{(\gamma_n-\gamma_{n-1})(\gamma_n-\delta_{n-1})}-\frac{\gamma_n\prod_{i=1}^{5}(\delta_{n}-e_i)}{(\delta_n-\gamma_{n-1})(\delta_n-\delta_{n-1})}\right).
\end{eqnarray}

We hope that the overdetermined system of equations (\ref{dubrt3}), (\ref{dubrt4}), (\ref{Rj}) can be used for constructing explicit solutions to the Volterra chain.
When $ m = 1 $, this is done below. However in generic case, the problem of solving these equations needs further investigation.

\section{Particular solutions to the Volterra chain.}

In this section, in a particular case we illustrate the application of the scheme above.  

\subsection{Construction of soliton solutions.}

Let us concentrate on the simplest nontrivial case $m=1$ by taking
\begin{equation}\label{parametrization_meq1}
P_{n}(\lambda)=\lambda-\gamma_n,\quad c(\lambda)=-\prod_{i=1}^{3}(\lambda-e_i).
\end{equation}
Now a system of equations (\ref{paramLaxt}), (\ref{paramLaxn}),  (\ref{un}) turns into
\begin{eqnarray}\label{paramLaxt_meq1}
\dot{\gamma}_n=u_n(\gamma_{n+1}-\gamma_{n-1}),
\end{eqnarray}
\begin{eqnarray}\label{paramLaxn_meq1}
u_n(2\lambda-\gamma_{n+1}-\gamma_{n})(2\lambda-\gamma_{n}-\gamma_{n-1})=\lambda(\lambda-\gamma_{n})^2- \prod_{i=1}^{3}(\lambda-e_i),
\end{eqnarray}
\begin{equation}\label{un_meq1}
u_{n}=-\frac{1}{2}\gamma_{n}+\frac{1}{4}\sum_{i=1}^{3}e_i.
\end{equation}
Similarly formulae (\ref{dubrn}), (\ref{dubrt1}) read as
\begin{equation}\label{dubrn_meq1}
(\gamma_{n}-\gamma_{n+1})(\gamma_{n}-\gamma_{n-1})=R(\gamma_{n}),
\end{equation}
\begin{eqnarray}\label{dubrt_meq1}
\dot{\gamma}_{n}=\left(-\frac{1}{2}\gamma_{n}+\frac{1}{4}\sum_{i=1}^{3}e_i\right)(\gamma_{n+1}-\gamma_{n-1}),
\end{eqnarray}
where
\begin{equation}
R(\gamma_{n})=-\frac{\prod_{i=1}^{3}(\gamma_{n}-e_i)}{-\frac{1}{2}\gamma_{n}+\frac{1}{4}\sum_{i=1}^{3}e_i}.
\end{equation}

{\bf Proposition 4.} The overdetermined system of the equations (\ref{dubrn_meq1}), (\ref{dubrt_meq1}) is compatible.

The proposition is proved by a direct computation. 

{\bf Corollary of Proposition 4}. A common solution of (\ref{dubrn_meq1}), (\ref{dubrt_meq1}) exists, it produces due to (\ref{un_meq1}) a particular solution to the Volterra chain. 

Obviously for the fixed value of $n$ the variables $\gamma_{n}$, $\gamma_{n-1}$ are found from the following system of the ordinary differential equations:
\begin{eqnarray}\nonumber
\dot{\gamma}_{n}=\left(-\frac{1}{2}\gamma_{n}+\frac{1}{4}\sum_{i=1}^{3}e_i\right)\left(\gamma_{n-1}-\gamma_{n}-\frac{R(\gamma_{n})}{\gamma_{n-1}-\gamma_{n}}\right),
\end{eqnarray}
\begin{eqnarray}\nonumber
\dot{\gamma}_{n-1}=\left(-\frac{1}{2}\gamma_{n}+\frac{1}{4}\sum_{i=1}^{3}e_i\right)\left(\gamma_{n}-\gamma_{n-1}-\frac{R(\gamma_{n-1})}{\gamma_{n-1}-\gamma_{n}}\right).
\end{eqnarray}
The discrete equation (\ref{dubrn_meq1}) specifies the dependence on the discrete variable $n$. The order of (\ref{dubrn_meq1}) is easily reduced since it admits an integral of motion:
\begin{equation}\label{dubrn2_meq1}
u_n(\gamma_{n}+\gamma_{n+1})(\gamma_{n}+\gamma_{n-1})=e_1e_2e_3
\end{equation}
that evidently follows from (\ref{paramLaxn_meq1}) with $\lambda=0$. 
Actually (\ref{dubrn_meq1}) provides an example of the discrete integrable map admitting a symmetry (\ref{dubrt_meq1}) (about discrete maps see \cite{Veselov2}, \cite{Nijhoff}, \cite{Gubbiotti} and references therein).  After some elementary transformations we get
\begin{eqnarray}\label{eq1_Gamma1}
\gamma_{n+1}=&-\frac{\gamma^2_{n}-(e_1+e_2+e_3)\gamma_{n}+e_1e_2+e_1e_3+e_2e_3}{2\gamma_{n}-e_1-e_2-e_3}\nonumber\\
&-\frac{S}{2\gamma_{n}-e_1-e_2-e_3},
\end{eqnarray}
\begin{eqnarray}\label{eq1_Gammat}
\gamma_{n,t}=\frac{S}{2},
\end{eqnarray}
where
\begin{eqnarray}\nonumber
S=\sqrt{(\gamma^2_{n}-e_1e_2-e_1e_3-e_2e_3)^2+4e_1e_2e_3(2\gamma_{n}-e_1-e_2-e_3)}.
\end{eqnarray}

It is easily checked that the overdetermined system (\ref{eq1_Gamma1}), (\ref{eq1_Gammat}) is consistent. Our goal now is to find a solution of the system and then due to the formula (\ref{un_meq1}) construct the corresponding solution of the Volterra chain.

Let us first consider the degenerate case when $e_3=e_2$. We fix the branch of the root $S$ assuming that for $\gamma\mapsto\infty$ its value is $\gamma^2$ and then rewrite the formulas (\ref{eq1_Gamma1}), (\ref{eq1_Gammat}) as follows
\begin{eqnarray}\label{gammat_solitoncase}
\gamma_{n,t}=\frac{\varepsilon(n)}{2}(e2-\gamma_{n})\sqrt{\gamma^2_{n}+2e_2\gamma_{n}+e_2^2-4e_1e_2},
\end{eqnarray}
\begin{eqnarray}\label{gammanp1_solitoncase}
\gamma_{n+1}=&-\frac{\gamma^2_{n}-(2e_2+e_1)\gamma_{n}+e_2^2+2e_1e_2}{2\gamma_{n}-e_1-2e_2}\nonumber\\
&+\frac{\varepsilon(n)(\gamma_{n}-e2)\sqrt{\gamma^2_{n}+2e_2\gamma_{n}+e_2^2-4e_1e_2}}{2\gamma_{n}-e_1-2e_2}.
\end{eqnarray}

It is convenient to return to the field variable $u_n(t)$ in the equations (\ref{gammat_solitoncase}), (\ref{gammanp1_solitoncase}). Due to (\ref{un_meq1}) we obtain:
\begin{eqnarray}\label{up1}
u_{n+1}&=\frac{8e_1u_n+16e_2u_n+e_1^2-4e_1e_2-16u_n^2}{32u_n}+\nonumber\\
& \, \frac{\varepsilon(n)(e_1-4u_n)\sqrt{16u_n^2-8e_1u_n-32e_2u_n+e_1^2-8e_1e_2+16e_2^2}}{32u_n},
\end{eqnarray}
\begin{eqnarray}\label{ut}
u_{n,t}=\frac{\varepsilon(n)(e_1-4u_n)\sqrt{16u_n^2-8e_1u_n-32e_2u_n+e_1^2-8e_1e_2+16e_2^2}}{16}.
\end{eqnarray}
We find in a standard way the solution of (\ref{ut}):
\begin{eqnarray}\label{u_ob}
u_n(t)=\frac{(e_2^2-w^2)\left(\frac{e_2-w}{e_2+w}e^{w(c(n)+\varepsilon(n)t)}-1\right)\left(\frac{(e_2+w)^2}{(e_2-w)^2}e^{w(c(n)+\varepsilon(n)t)}-1\right)}{4e_2\left(e^{w(c(n)+\varepsilon(n)t)}-1\right)\left(\frac{e_2+w}{e_2-w}e^{w(c(n)+\varepsilon(n)t)}-1\right)},
\end{eqnarray}
where $\omega=\pm \sqrt{e_2(e_2-e_1)}$, it is supposed that $e_2(e_2-e_1)>0$ and $c(n)$ is a function of n. Let us determine the explicit form of $c(n)$ by using the equation (\ref{up1}). 

Since it depends on the choice of $\varepsilon(n)$ we consider separately all of the possible cases $\varepsilon=1$, $\varepsilon=-1$, $\varepsilon=(-1)^n$.

Assume that $\varepsilon=1$ and substitute (\ref{u_ob}) into equation (\ref{up1}). Then we get immediately that $c(n)$ solves one of the following equations
\begin{eqnarray*}
(e_2-w)e^{wc(n+1)}-(e_2+w)e^{wc(n)}=0, \label{eq1_cn}\\
(e_2+w)e^{wc(n+1)}-(e_2-w)e^{wc(n)}=0, \label{eq2_cn}
\end{eqnarray*}
which obviously imply:
\begin{eqnarray}
c(n)=\frac{1}{w} \left(n\,ln\left(\frac{e_2+w}{e_2-w}\right)+c_1+i\pi\right),\label{cn_11}\\
c(n)=\frac{1}{w} \left(n\,ln\left(\frac{e_2-w}{e_2+w}\right)+c_2+i\pi\right),\label{cn_12}
\end{eqnarray}
respectively. Here we request that $c_1$, $c_2$ are arbitrary constants, we put summand $i\pi$ in order to change the sign before the exponentials in (\ref{u_ob}). Direct computation convinces us that $u_n(t)$ defined in (\ref{u_ob}) solves the Volterra chain if $c(n)$ is given by (\ref{cn_11}):
\begin{eqnarray}\label{u_cn11}
u_n(t)=\frac{(e_2^2-w^2)\left(e^{(n-1)\theta+wt+c_1}+1\right)\left(e^{(n+2)\theta+wt+c_1}+1\right)}{4e_2\left(e^{n\theta+wt+c_1}+1\right)\left(e^{(n+1)\theta+wt+c_1}+1\right)},
\end{eqnarray}
where $\theta=ln\left(\frac{e_2+w}{e_2-w}\right)$.

Obviously these solutions coincide with Manakov's solutions found in \cite{Manakov} (see also \cite{Novikovbook}) up to notations. The cases $\varepsilon=-1,$ $\varepsilon=(-1)^{n}$ lead to the same solution (\ref{u_cn11}).

\subsection{Construction of elliptic solutions.}

Here we focus on a pair of compatible equations (\ref{eq1_Gamma1}) and (\ref{eq1_Gammat}). For the simplicity we introduce notations $\lambda_1=-e_1-e_2-e_3$, $\lambda_2=e_1e_2+e_1e_3+e_2e_3$ and $\lambda_3=-e_1e_2e_3$ and rewrite the equations as
\begin{eqnarray}\label{eq3_Gamma1}
\gamma_{n+1}=&-\frac{\gamma^2_n+\lambda_1\gamma_n+\lambda_2}{2\gamma_n+\lambda_1}-\frac{\sqrt{\gamma^4_n-2\lambda_2\gamma^2_n-8\lambda_3\gamma_n+\lambda^2_2-4\lambda_1\lambda_3}}{2\gamma_n+\lambda_1},
\end{eqnarray}
\begin{eqnarray}\label{eq3_Gammat}
\gamma_{n,t}=\frac{\sqrt{\gamma^4_n-2\lambda_2\gamma^2_n-8\lambda_3\gamma_n+\lambda^2_2-4\lambda_1\lambda_3}}{2}.
\end{eqnarray}

To solve the equation (\ref{eq3_Gammat}), we first convert the polynomial 
\begin{equation}\label{G} 
G(\gamma_n):=\frac{1}{4}\gamma_n^4-\frac{1}{2}\lambda_2\gamma^2_n+2\lambda_3\gamma_n+ \frac{\lambda^2_2-4\lambda_1\lambda_3}{4}=y_n^2
\end{equation}
into the Weierstrass normal form by applying the following fractionally rational transformation (for more details see the book \cite{Bateman})
\begin{equation}\label{transform} 
\gamma_n= \frac{2\alpha_0\xi_n-\alpha_0 A_2-2A_1}{2\xi_n-A_2},\quad y_n=\frac{A_1\eta_n}{(\xi_n-\frac{1}{2}A_2)^2},
\end{equation}
where $A_1=-\frac{1}{4}\alpha_0^3+\frac{1}{4}\alpha_0\lambda_2+\frac{1}{2}\lambda_3$, $A_2=\frac{1}{4}\alpha_0^2-\frac{1}{12}\lambda_2$ and $\alpha_0$ is a root of the polynomial $G(\gamma_n)$.

In terms of the new variables $\xi_n$, $\eta_n$ we get an elliptic curve
\begin{equation}\label{eta} 
4\xi_n^3-g_2\xi_n-g_3=\eta_n^2,
\end{equation}
where $g_2=\frac{1}{12}\lambda^2_2-\frac{1}{4}\lambda_1\lambda_3$,  $g_3=-\frac{1}{216}\lambda^3_2-\frac{1}{16}\lambda_3^2+ \frac{1}{48}\lambda_1\lambda_2\lambda_3$. 
As a result of the transformation (\ref{transform}) equations (\ref{eq3_Gamma1}) and  (\ref{eq3_Gammat}) are reduced to the form
\begin{equation}\label{weier} 
\xi_{n,t}=\sqrt{4\xi_n^3-g_2\xi_n-g_3}
\end{equation}
and
\begin{eqnarray}\label{xinp1} 
\xi_{n+1}=\frac{\lambda_2}{12}+\frac{1}{4}\frac{3\lambda_1\lambda_3}{12\xi_n-\lambda_2}+\frac{1}{2}\frac{9\lambda_3^2}{(12\xi_n-\lambda_2)^2}\nonumber\\
\qquad \qquad \qquad -\frac{18\lambda_3\sqrt{4\xi_n^3-g_2\xi_n-g_3}}{(12\xi_n-\lambda_2)^2}.
\end{eqnarray}

Since the point transformation preserves the compatibility condition equations (\ref{weier}), (\ref{xinp1}) are also compatible. Now our goal is to construct common general solution to (\ref{weier}), which is obviously expressed in terms of the Weierstrass $\wp$ function
\begin{equation}\label{xin} 
\xi_n(t)=\wp(t+c(n)).
\end{equation}
Actually we have to find the explicit form of the function $c(n)$ in the formula (\ref{xin}) such that it solves (\ref{xinp1}) as well.

{\bf Proposition 5.} Function $\xi_n(t)$ can be taken as
\begin{eqnarray}\label{xinwpeps1}
\xi_n(t)=\wp(t+v n+\beta).
\end{eqnarray}

{\bf Proof.} We rewrite formula (\ref{xinp1}) in the form
\begin{eqnarray}\label{addth_xi}
\xi_{n+1}=\frac{1}{4}\left[\frac{\sqrt{4\xi_n^3-g_2\xi_n-g_3}-\frac{\lambda_3}{4}}{\xi_n-\frac{\lambda_2}{12}}\right]^2-\xi_n-\frac{\lambda_2}{12}.
\end{eqnarray}
It is remarkable that (\ref{addth_xi}) looks very similar to the addition theorem for the Weierstrass $\wp$ function
\begin{eqnarray}\label{addth}
\wp(u+v)=\frac{1}{4}\left[\frac{\wp'(u)-\wp'(v)}{\wp(u)-\wp(v)}\right]^2-\wp(u)-\wp(v),
\end{eqnarray}
where $\wp'(u)$ is the derivative of the function $\wp(u)$ (see \cite{Bateman}). 

We show that for an appropriate choice of the parameters these two equations coincide completely. In (\ref{addth}) we set $c(n)=vn+\beta$, $u=t+v n+\beta$, where $v$ is found from the equation $\wp(v)=\frac{\lambda_2}{12}$. It is easily verified that $\wp'(v)$ can be chosen in such a way $\wp'(v)=\sqrt{4\left(\frac{\lambda_2}{12}\right)^3-g_2\frac{\lambda_2}{12}-g_3}=\frac{\lambda_3}{4}$. Then actually we get coincidence of the equations (\ref{addth_xi}), (\ref{addth}) that implies that function (\ref{xinwpeps1}) solves equations (\ref{weier}) and (\ref{xinp1}). That completes the proof.

Now it remains to express $\gamma_n(t)$ through $\xi_n(t)$ and then write down a solution $u_n(t)$ of the Volterra chain.

For known $\xi_n(t)$ we can recover $\gamma_n(t)$ due to the M$\ddot{o}$bius transformation (\ref{transform}) 
\begin{eqnarray}
\gamma_n(t)=\frac{2\alpha_0\wp(z)-2A_1-\alpha_0A_2}{2\wp(z)-A_2},\label{transformMobius}
\end{eqnarray}
where $z=t+v n+\beta$. Then solution of the Volterra chain is given by (\ref{un_meq1}), i.e.
\begin{eqnarray}
u_n(t)=\frac{r_2\wp(z)+r_1}{\wp(z)+r_3},\label{solution_u}
\end{eqnarray}
where
\begin{eqnarray*}
&r_1=\frac{1}{2}A_1+\left(\frac{1}{4}\alpha_0+\frac{1}{8}\lambda_1\right)A_2,\\
&r_2=-\frac{1}{2}\alpha_0-\frac{1}{4}\lambda_1, \quad r_3=-\frac{1}{2}A_2.
\end{eqnarray*}

In order to transform (\ref{solution_u}) to an accepted form we express $u_n(t)$ in terms of the Weierstrass $\sigma$ function. According to the well known theorem \cite{Bateman} we have
\begin{eqnarray}
u_n(t)=C\frac{\sigma(z-\mu)\sigma(z+\mu)}{\sigma(z-\theta)\sigma(z+\theta)},\label{solution_usigma}
\end{eqnarray}
where $\mu$ and $\theta$ are found from the equations
\begin{eqnarray}
\wp(\mu)=-\frac{r_1}{r_2}, \quad \wp(\theta)=-r_3. \label{solution_muteta}
\end{eqnarray}
We find $C$ due to the fact that $\wp(z)$ has a pole at the point $z=0$:
\begin{eqnarray}
C=r_2\frac{\sigma^2(\theta)}{\sigma^2(\mu)}. \label{C}
\end{eqnarray}
Let us apply to (\ref{solution_usigma}) the point transformation $u\rightarrow r_2 u$, $t\rightarrow r_2 t$ preserving the Volterra chain and bring the found $u_n(t)$ to the form (cf. \cite{Veselov})
\begin{eqnarray}
u_n(t)=\frac{\sigma^2(\theta)}{\sigma^2(\mu)}\frac{\sigma(r_2t+vn+\beta-\mu)\sigma(r_2t+vn+\beta+\mu)}{\sigma(r_2t+vn+\beta-\theta)\sigma(r_2t+vn+\beta+\theta)}.\label{ves-solution_usigma}
\end{eqnarray}

\section*{Conclusions} 

The article proposes an alternative way for constructing exact solutions for integrable models based on the concept of generalized invariant manifolds. We have illustrated the scheme with an example of the Volterra chain. The first step is to derive an appropriate generalized invariant manifold to the chain, containing at least two constant parameters that are not removed by a point transformation. Actually such a manifold determines a nonlinear Lax pair. Next we assume that this Lax pair has a solution polynomially depending  on one of the parameters $\lambda$. We also require that the other parameter is a polynomial on $\lambda$, as well. It turned out that under these requirements the solution of the nonlinear Lax pair satisfies to an overdetermined system of differential and difference equations. In the case of small degree polynomials, in more details studied in the article, we arrive at a pair of consistent equations. One of the equations is differential, it defines an elliptic function and the other one is difference, it coincides with the addition theorem for that elliptic function. In the case of arbitrary $m$ system of equations (\ref{dubrt3}), (\ref{dubrt4}) with additional constraint (\ref{Rj}) are derived. However, the problem of solving these equations by the methods developed in \cite{DubrovinMatveevNovikov} (see also \cite{Veselov2}, \cite{Smirnov}, \cite{Bobenko}) needs further investigation.

\subsection*{Acknowledgements}
We thank M.V.Pavlov for his interest to the work.


\begin{thebibliography}{20}

\bibitem{HabKhaPo}  Habibullin I T, Khakimova A R and Poptsova M N 2016 On a method for constructing the Lax pairs for nonlinear integrable equations {\it J.
Phys. A: Math. Theor.} {\bf 49} 035202

\bibitem{HabKhaTMP17} Habibullin I T and Khakimova A R 2017 Invariant manifolds and the Lax pairs for integrable nonlinear lattices {\it Theor. Math. Phys.} {\bf 191} 793-810    

\bibitem{HabKhaJPA17} Habibullin I T and Khakimova A R 2017 On a method for constructing the Lax pairs for integrable models via a quadratic ansatz {\it J.
Phys. A: Math. Theor.} {\bf 50} 305206

\bibitem{HabKhaTMP18} Habibullin I T and Khakimova A R 2018 A direct algorithm for constructing recursion operators and Lax pairs for integrable models {\it Theor. Math. Phys.} {\bf 196} 1200-1216

\bibitem{KhaUMJ18} Khakimova A R 2018 On description of generalized invariant manifolds for nonlinear equations {\it Ufa Math. J.} {\bf 10} 106-116

\bibitem{Volterra} Volterra V 1931 {\it Le$c_{,}$ons sur la th$\acute{e}$orie math$\acute{e}$matique de la lutte pour la vie} (Paris: Gauthier-Villars)  

\bibitem{Manakov} Manakov S V 1974 Complete integrability and stochastization of discrete dynamical systems {\it Zh. Eksp. Teor. Fiz} {\bf 67} 543-555

\bibitem{KacvanMoerbeke} Kac M and van Moerbeke P 1975 A complete solution of the periodic Toda problem {\it Proc. Nat. Acad. Sci. USA } {\bf 72} 2879-2880

\bibitem{Novikovbook} Novikov S, Manakov S V, Pitaevskii L P and Zakharov V E 1984 {\it Theory of Solitons: The Inverse Scattering Method} (Springer Science, Business Media)

\bibitem{Veselov} Veselov A P 1987 Integration of the stationary problem for a classical spin chain {\it Theor. Math. Phys.} {\bf 71} 446-450

\bibitem{Ver} Vereshchagin V L 1988 Hamiltonian structure of averaged difference systems {\it Math. Notes} {\bf 44} 798-805

\bibitem{Solomon} Alber S J 1991 Associated integrable systems {\it  J. Math. Phys.} {\bf 32} 916-922

\bibitem{DubrovinMatveevNovikov} Dubrovin B A, Matveev V B and Novikov S P 1976 Non-linear equations of Korteweg-de Vries type, finite-zone linear operators, and Abelian varieties {\it Russian Math. Surveys} {\bf 31} 59-146

\bibitem{Krichever77}  Krichever I M 1977 Integration of nonlinear equations by methods of algebraic geometry 
{\it Funktsional. Anal. i Prilozhen.} {\bf 11(1)} 15-31

\bibitem{ItsKitaevFokas} Its A R, Kitaev A V and Fokas A S 1990 The isomonodromy approach in the theory of two-dimensional quantum gravitation {\it Russian Math. Surveys} {\bf 45} 155-157

\bibitem{FokasItsKitaev} Fokas A S, Its A R and Kitaev A V 1991 Discrete Painleve equations and their appearance in quantum gravity {\it Commun. Math. Phys.} {\bf 142} 313-344

\bibitem{AdlerShabat1} Adler V E and Shabat A B 2018 Volterra chain and Catalan numbers {\it JETP Letters} {\bf 108} 825-828

\bibitem{AdlerShabat2} Adler V E and Shabat A B 2019 Some exact solutions of the Volterra lattice {\it Theor. Math. Phys.} {\bf 201} 1442-1456

\bibitem{Nijhoff1} Nijhoff F W 2000 Discrete Dubrovin equations and separation of variables for discrete systems {\it Chaos Solitons Fractals} {\bf 11(1)} 19-28

\bibitem{ZhangZY} Zhang Z Y 2019 An upper order bound of the invariant manifold in Lax pairs of a nonlinear evolution partial differential equation {\it J. Phys. A: Math. Theor.} {\bf 52} 265202

\bibitem{Calogero} Ahmed S, Bruschi M, Calogero F, Olshanetsky M A and Perelomov A M 1979 Properties of the zeros of the classical polynomials and of the Bessel functions {\it Il Nuovo Cimento B}  {\bf 49(2)} 173-199

\bibitem{Veselov2} Veselov A P 1991 Integrable maps {\it Russian Math. Surveys} {\bf 46} 1-51  

\bibitem{Nijhoff} Atkinson J, Lobb S B and Nijhoff F W 2012 An integrable multicomponent quad-equation and its Lagrangian formulation {\it Theor. Math. Phys.} {\bf 173} 1644-1653

\bibitem{Gubbiotti} Gubbiotti G 2019 Lagrangians and integrability for additive fourth-order difference equations (arXiv:1910.11458)

\bibitem{Bateman} Bateman H and Erdelyi A 1955 {\it Higher transcendental functions} (McGraw-Hill Book Company, New York) {\bf 3}

\bibitem{Smirnov} Golovachev G M and Smirnov A O 2010 On spectral curve for functional-difference Shrodinger equation {\it J. Math. Sci. (N. Y.)} {\bf 168} 820-828

\bibitem{Bobenko} Bikbaev R F, Bobenko A I and Its A R 2014 Landau-Lifshitz equation, uniaxial anisotropy case: Theory of exact solutions {\it Theor. Math. Phys.} {\bf 178} 143-193

\end{thebibliography}
\end{document}